\documentclass[usenatbib]{mn2e}

\usepackage{graphicx}
\usepackage{amssymb}
\newcommand{\text}[1]{\quad\mbox{#1}\quad}

\newcommand{\be}{\begin{equation}}
\newcommand{\ee}{\end{equation}}

\newcommand{\fracb}[2]{\left(\frac{#1}{#2}\right)}

\newcommand{\mean}[1]{\langle{#1}\rangle}

\begin{document}

\title[ Magnetic dissipation in the Crab nebula]
{Magnetic dissipation in the Crab nebula}

\author[Komissarov S.S.]
{
Serguei S.~Komissarov
\\
Department of Applied Mathematics, The University of Leeds, Leeds, LS2 9GT, UK\\
E-mail: serguei@maths.leeds.ac.uk
}

\date{Received/Accepted}
\maketitle

\begin{abstract}
Magnetic dissipation is frequently invoked as a way of powering 
the observed emission of relativistic flows in Gamma Ray Bursts and 
Active Galactic Nuclei. Pulsar Wind Nebulae provide closer to home 
cosmic laboratories which can be used to test the hypothesis.   
To this end, we reanalyze the observational data on the spindown 
power of the Crab pulsar, energetics of the Crab nebula, and its 
magnetic field. We show that unless the magnetic inclination angle 
of the Crab pulsar is very close to 90 degrees the overall 
magnetization of the striped wind after total dissipation of 
its stripes is significantly higher than that deduced in the 
Kennel-Coroniti model and recent axisymmetric simulations of 
Pulsar Wind Nebulae. On the other hand, higher wind magnetization  
is in conflict with the observed low magnetic field of the Crab nebula, 
unless it is subject to efficient dissipation inside the nebula as well. 
For the likely inclination angle of 45 degrees the data 
require magnetic dissipation on the timescale about  
80 years, which is short compared to the life-time of the nebula but 
long compared to the time scale of Crab's gamma-ray flares.          
\end{abstract}


%sssssssssssssssssssssssssssssssssssssssss
\section{Introduction}
\label{sec:intr}
%sssssssssssssssssssssssssssssssssssssssss

Magnetic fields are often invoked in models of the relativistic jet
production by central engines of Active Galactic Nuclei (AGN) and
Gamma Ray Bursts (GRB). In these theories the jets are
Poynting-dominated at the origin, with the magnetization parameter
$\sigma=B^2/4\pi\rho c^2\gg 1$.  This is different from the earlier
essentially hydrodynamic, low $\sigma$, models of relativistic jets in
one important aspect. Even strong, high Mach number shocks, in high
$\sigma$ plasma are weakly dissipative compared to their low $\sigma$
counterparts \citep[e.g.][]{KC84a,K12}. Moreover, PIC simulations show
that the acceleration of nonthermal particles may also be problematic
at such shocks \citep{SS09,SS11a}. This suggests that either the
Poynting flux is first converted into the bulk motion kinetic energy
via ideal MHD mechanism \citep[e.g.][]{VK04,KVKB09,L10}, which is then
dissipated at shocks, or the magnetic energy is converted directly into
the energy of emitting particles via magnetic dissipation, which
accompanies magnetic reconnection events
\citep[e.g.][]{DS02,LB03,ZY11,G11,MU12}.  In fact, the magnetic
dissipation can facilitate bulk acceleration of jets as well 
\citep[e.g.][]{DS02}.

While AGN and GRBs are very distant sources, which makes their
observational studies rather difficult, there exist objects much
``closer to home'' which share similar properties, the Pulsar Wind
Nebulae (PWN). They are powered by relativistic winds from neutron
stars and these winds are also expected to be Poynting-dominated at
their base \citep[see ][and references therein]{Ar12}.  In particular,
the Crab nebula is one of the brightest sources of nonthermal emission 
in the sky throughout the whole observational range of photon energies. 
Its large angular size (of seven arc minutes), ensures that its spatial 
structure is well resolved and its relatively small linear size 
(of several light years) allows direct observations of not only  
its small-scale structural variability but also its overall 
dynamics.         
Because the Crab nebula is such an easy object to observe it has 
been studied with the level of detail which may never be reached 
in observations of AGN and GRB jets, and it is rightly considered 
as a testbed of relativistic astrophysics.  
  
The early attempts to built a theoretical model of the Crab nebula
using the ideal relativistic MHD approximation resulted in a
paradoxical conclusion that the pulsar wind has to have
$\sigma\sim10^{-3}$ near its termination shock \citep{RG74,KC84a,EC87,BL92}. 
A slightly higher magnetization, $\sigma\sim10^{-2}$, was later
suggested by axisymmetric numerical simulations
\citep{KL03,LDZ04,B05}, although no proper study of this issue has
been carried out.  The key property of these analytical and numerical
solutions is their purely toroidal magnetic field. The strong hoop stress of
such field creates excessive axial compression of the nebula in
solutions with higher $\sigma$ and pushes the termination shock too
close to the pulsar in the Kennel-Coroniti model, in conflict with the
observations. On the other hand, the ideal relativistic MHD
acceleration of uncollimated wind-like flows is known to be very
inefficient, leaving such flows Poynting-dominated on the
astrophysically relevant scales \citep[e.g.][]{L11,K11}.  This
striking conflict is known as the $\sigma$-problem.

Attempts have been made to see if $\sigma$ can be reduced via magnetic
dissipation in the so-call striped zone of pulsar winds, where the
magnetic field changes it polarity on the length scale $\lambda_p
=cP$, where $P$ is the pulsar period \citep{C90,LK01}.  The
dissipation is accompanied by the wind acceleration via conversion of
the thermal energy into the bulk kinetic energy of the flow during its
adiabatic expansion.  Unfortunately, for the wind of the Crab pulsar
the dissipation length scale significantly exceeds the radius of the
wind termination shock, thus making this mechanism inefficient
\citep{LK01}.
  
\citet{L03b} has demonstrated that the energy associated with the
alternating component of magnetic field of the striped wind can be
rapidly dissipated at the termination shock itself, where the
characteristic Larmor radius of shock-heated plasma exceeds the
wavelength of magnetic stripes.  His solution of the shock equations, which
accounts for the ``erasing'' of stripes, shows that the post-shock
flow is the same as it would be if the dissipation had already been
fully completed in the wind.  \citet{SS11b} have used 3D PIC
simulations to study the magnetic dissipation and particle
acceleration at the termination shock of the striped wind numerically
and concluded that efficient magnetic dissipation occurs even when the
Larmor radius remains below the stripes wavelength, via rapid
development of the tearing mode instability and magnetic reconnection
in the post-shock flow.

One way or another, this dissipation occurs only in the striped zone
and only the alternating component of magnetic field dissipates.
Outside of the striped zone, around the poles, the pulsar wind
$\sigma$ remains unaffected by this dissipation and hence very high.
As the result, the overall magnetization of plasma injected into the 
nebula can be much higher than that of the Kennel-Coroniti model,  
unless the striped zone spreads over almost the entire wind
\citep{C90}.  

\citet{L03a} argued that in the polar zone the wind $\sigma$ can be
reduced via the flow acceleration accompanying dissipation of fast
magnetosonic waves emitted by the pulsar into the polar zone.
However, it seems unlikely that the energy flux associated with these
waves can dominate the wind energetics in the polar zone. At least,
the 3D numerical simulations of pulsar winds (A.Spitkovsky, private
communication) show that their contribution is rather small.  Thus, we
do not expect $\sigma$ of the polar zone to be below unity.

An alternative solution to the $\sigma$ problem has been proposed by
\citet{B98}, who argued that the axial compression of the nebula can
be reduced via the current-driven kink instability, resulting in more
or less uniform total pressure distribution inside the nebula. This
would make the overall structure and dynamics of the nebula similar to
those in the models with particle-dominated ultra-relativistic pulsar wind.  
The recent computer simulations of the non-linear development of the kink
instability of relativistic z-pinch configurations support this
conclusion \citep{M09,M11}. In this scenario, PWN are supplied with
highly magnetized plasma, making magnetic dissipation a potentially
important process in their evolution and emission.

In this paper, we test whether the magnetic dissipation inside PWN is
consistent with the observations of the Crab nebula and its pulsar.
The main idea is very simple.  First, the timing observations of the
Crab pulsar allow us to estimate how much energy has being pumped into
the nebula.  Second, using the stripe wind model we can calculate how
much of this energy is supplied in the magnetic form. Third, a simple
dynamical model of the nebula expansion can be used to predict how
much magnetic energy is retained by the nebula after adiabatic losses.
Finally, the observations of the Crab nebula tell us how much magnetic
energy is actually in there and whether the magnetic dissipation is
actually required to make the ends meet.

%sssssssssssssssssssssssssssssssssssssssssssssssssssssssssssssssss
\section{Overall energetics of the Crab nebula} 
\label{sec:supply}
%sssssssssssssssssssssssssssssssssssssssssssssssssssssssssssssssss

In the simplest approximation, the spindown of pulsars is described by
the equation $\dot\Omega\propto-\Omega^n$, where $\Omega$ is the
pulsar angular frequency and $n$ is the so-called braking index. This
form of the spindown law originates from the magneto-dipole vacuum
radiation mechanism which gives $n=3$. Force-free (or magnetodynamic)
models of pulsar magnetospheres yield the same dependence on $\Omega$
\citep{S06,KC09}. Timing observations of pulsars allow to measure the 
braking index and it turns out to be noticeably lower compared to the value 
predicted by these simple models \citep{LPS93}. The reason for this 
discrepancy is not established yet, but the spindown law itself seems to 
be consistent with the observations and we will accept it in our calculations.
  
The solution to this equation is 

\be \Omega =
\Omega_0\left(1+\frac{t}{\tau}\right)^{-\frac{1}{n-1}}\, .
\label{Om-t}
\ee 
The corresponding spindown luminosity is 

\be L_{sp} =
-I\Omega\dot{\Omega} =
L_0\left(1+\frac{t}{\tau}\right)^{-\frac{n+1}{n-1}}\, ,
\label{L-t}
\ee 
where $\tau$ is called the spindown time \citep{RG74}.  From the
timing observations of the Crab pulsar, $n=2.51$ and $\tau\simeq
703\,$yr \citep{LPS93}. 
For the usually accepted moment of inertia of neutron stars 
$I=10^{45}\mbox{g}\,\mbox{cm}^2$, these measurements imply the current 
spindown power $L_{sp} \simeq 4.6\times10^{38}\mbox{erg/s}$ and the initial
power $L_0\simeq 3.3\times10^{39}\mbox{erg/s}$.  The corresponding total 
extracted rotational energy of the Crab pulsar is 

\be
   E = L_0\tau\frac{n-1}{2}
  \left(1- \left(1+\frac{t}{\tau}\right)^{-\frac{2}{n-1}}\right) 
  \simeq 3.7\times10^{49}\mbox{erg}\, , 
\label{E}
\ee
which is 67 percent of its initial rotational energy.  
The integrated radiative luminosity of the Crab nebula 
$L_n\simeq 1.3\times10^{38}\mbox{erg/s}$ \citep{H08} is significantly below 
$L_{sp}$. Thus, a large fraction of $E$ is converted into the kinetic 
energy of the supernova shell and the internal energy of the PWN,  
the actual proportion being dependent on the dynamic evolution of the 
nebula.

%fffffffffffffffffffffffffffffffffffffffffffffffffffffffffffffffffff
\begin{figure*}
\includegraphics[width=58mm]{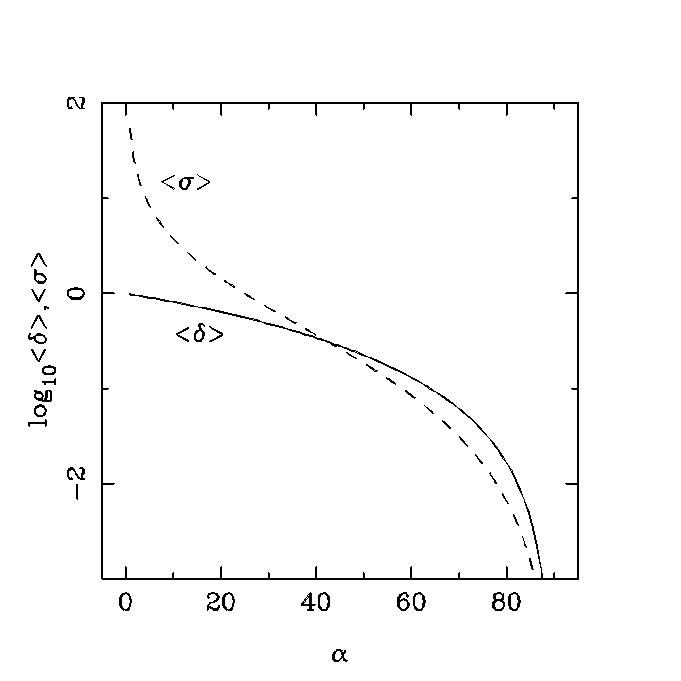}
\includegraphics[width=58mm]{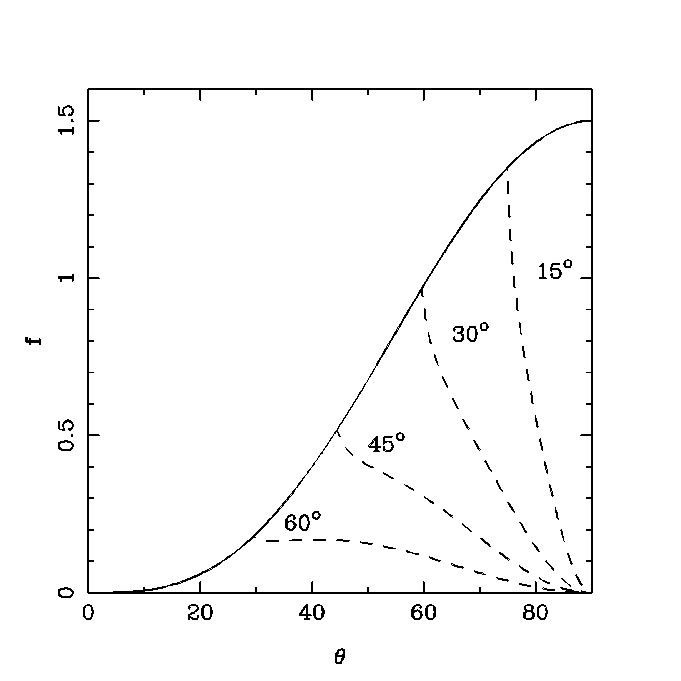}
\includegraphics[width=58mm]{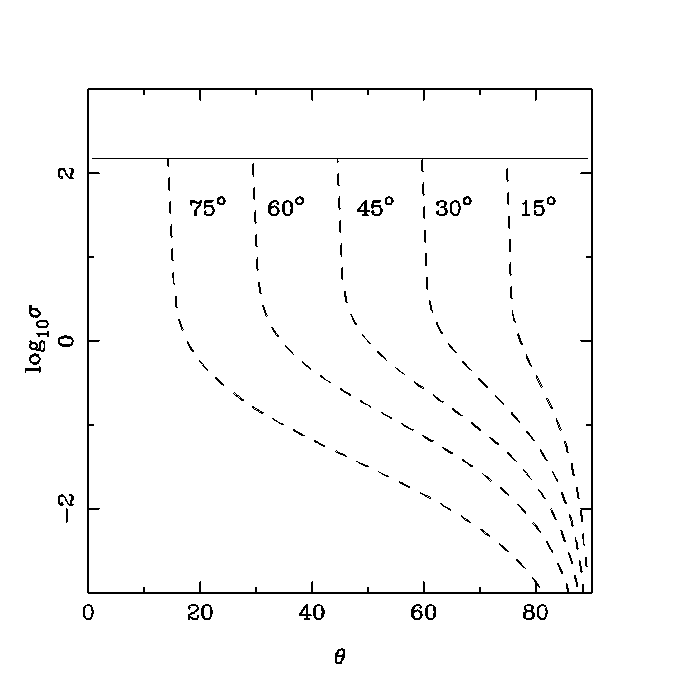}
\caption{
{\it Left panel:} 
Mean magnetization $\mean{\sigma_\alpha}$ of the pulsar wind after dissipation 
of its stripes and the mean fraction $\mean{\delta_\alpha}$ of 
magnetic magnetic energy injected into PWN as functions of the pulsar 
magnetic inclination angle.
{\it Middle panel:} 
The distribution of magnetic energy injected into PWN over the 
polar angle, 
$f_\alpha(\theta) = (3/2)\delta_\alpha(\theta)\sin^3\theta$, for 
the magnetic inclination angle $\alpha=0^o$ 
(solid line) $15^o,30^o,45^o,60^o$ (dashed lines).   
{\it Right panel:} 
The magnetization of the striped wind after dissipation of its stripes 
as a function of the polar angle for the magnetic inclination angle 
$\alpha=0^o$ (solid line), $15^o,30^o,45^o,60^o,75^o$ (dashed lines).    
}
\label{fig1}
\end{figure*}
%fffffffffffffffffffffffffffffffffffffffffffffffffffffffffffffffffff

Since $\tau$ is comparable to the current age of the nebula, its global 
dynamics cannot be described by self-similar models. This forces us to 
make a number of strong simplifying assumptions in order to render the 
problem treatable. First, we assume that the nebula is uniform. Second, that     
the  magnetic field becomes randomized, via development of instabilities,  
and behaves as gas with ultrarelativistic ratio of specific heats 
$\Gamma=4/3$. In this case, the internal energy of PWN is $E_n=3pV$, 
where $p$ is the PWN uniform total pressure and $V$ is its volume. 
Third, that the nebula expands with constant speed, and hence $V\propto t^3$, 
which is supported by the rather slow observed acceleration of the 
nebula \citep{T68,WM77,BKHW91}.    
Finally, we will ignore its radiative cooling\footnote{The radiative 
cooling would reduce the energy of relativistic particles $E_e$, making it 
even smaller compared to the magnetic energy $E_m$ than in our calculation. 
Ultimately, this would make the case for magnetic dissipation even stronger.}. 
Under these assumptions the evolution of the internal energy of the nebula 
is described by the equation

\be 
\dot{E_n}= L_{sp} - \frac{E_n}{t}, 
\ee 
where the last term describes adiabatic cooling. Given the expression
(\ref{L-t}) for $L_{sp}$, the initial condition $E_{n}(0)=0$, and
assuming $n<3$, we find the solution to this equation 

\be 
E_n=
\frac{L_0\tau}{a^2-3a+2} \frac{1}{x} \left( 1-
\frac{(a-1)x^2+ax+1}{(x+1)^a}\right)\, ,
\label{E-n}
\ee
where $a=(n+1)/(n-1)$ and $x=t/\tau$. For the parameters of the
Crab pulsar, this yields $E_n\simeq 1.3\times10^{49}\mbox{erg}$. The
corresponding spindown energy converted into the kinetic energy of the
supernova shell is then $E_k \simeq 2.4\times 10^{49}\mbox{erg}$.
The observations indicate that the expansion velocity of the thermal filaments of 
the Crab nebula increased by 100-200 km/s during the nebula lifetime, 
which corresponds to increase of their kinetic energy  
by $E_k\simeq 10^{49}$erg \citep{H08}. This agrees rather well with the 
prediction of our model. 

The internal energy $E_n$ is distributed between the relativistic particles
and the magnetic field. The actual partition is dictated by the properties 
of the pulsar wind, which determine how much energy is injected into the nebula 
in the magnetic form, and by the interaction between these two components inside 
the nebula. This interaction can have a reversible form, via the Lorentz force, 
and an irreversible form, e.g. via magnetic reconnection, collisionless wave 
dumping and particle acceleration. If the magnetic field is indeed significantly 
randomized, as we have assumed above, and the Lorentz force is reduced to 
the magnetic pressure then the reversible interaction is likely to be weak.          
As a first approximation, we will assume that the irreversible interaction is 
also weak, in which case the energy distribution between particles and magnetic 
fields in the nebula equals to that immediately downstream of the termination 
shock. By comparing the outcome with the observational data, we will be able 
to say how bad this assumption is and to gauge the importance of magnetic 
dissipation.

%sssssssssssssssssssssssssssssssssssssssssssssssssssssssssssssssss
\section{The magnetic power of striped wind} 
\label{sec:SW}
%sssssssssssssssssssssssssssssssssssssssssssssssssssssssssssssssss

In order to estimate the fraction of the wind energy injected into the nebula 
in the magnetic form we will employ the split-monopole model by \citet{B99}
and the finding of \citet{L03b} that the overall effect of the stripes 
dissipation at the termination shock is equivalent to their dissipation upstream 
of the shock.

Let us denote the angle between the spin axis and the magnetic axis of the pulsar, 
the magnetic inclination angle, as $\alpha$, the angle between the rotation 
axis and selected streamline of the wind as $\theta$ , and the phase of the 
stripe wave as $\phi$, with 
$\phi=0$ corresponding to the middle of the stripe with positive (or
negative) $B_\phi$.  Then the phases separating the positive and
negative stripes are $\phi_\alpha(\theta)$ and
$2\pi-\phi_\alpha(\theta)$ where

$$ 
\cos\phi_\alpha(\theta) = -\cot(\alpha)\cot(\theta).
$$ 
The conservation of total magnetic flux corresponding to one
wavelength allows us to find the magnitude of magnetic field after 
complete dissipation of its stripes as

$$ 
B = B_0 \left\{
    \begin{array}{ll}
       |2\phi_\alpha(\theta)/\pi-1|, & \pi/2-\alpha<\theta<\pi/2\\ 1,
       & \theta\le\pi/2-\alpha
\end{array}
\right. ,
$$ 
where $B_0$ is the magnitude of the magnetic field of the striped
wind (In \citet{L03b}, $B$ is called the mean magnetic field of the
striped wind). In these calculations we assume that after completion of 
this dissipation, the
relic current sheets collapse following their adiabatic cooling.  The
fraction of wind power remaining in the form of Poynting flux
along the stream line with the polar angle $\theta$ is

\be 
\chi_\alpha(\theta) = \left\{
    \begin{array}{ll}
       (2\phi_\alpha(\theta)/\pi-1)^2, &
      \pi/2-\alpha<\theta<\pi/2\\ 1, & \theta\le\pi/2-\alpha
\end{array}
\right. .  
\ee 
Neglecting the small initial contribution of the bulk kinetic
energy to the wind power (due to the ideal MHD acceleration in the wind), 
the wind magnetization along the stream line after the dissipation is

$$ 
\sigma_\alpha(\theta) =
\frac{\chi_\alpha(\theta)}{1-\chi_\alpha(\theta)} \, .
$$ 
We define the mean magnetization of the wind,
$\mean{\sigma_\alpha}$, as the ratio of its total Poynting flux to its
total bulk kinetic energy flux. Since in the split monopole model the
energy flux density varies with $\theta$ like $\sin^2\theta$, \be
\mean{\sigma_\alpha} = \frac{\mean{\chi_\alpha}}{1-\mean{\chi_\alpha}}
\, , \ee where

$$ \mean{\chi_\alpha} = \int\limits_0^{\pi/2}
\chi_\alpha(\theta)\sin^3\theta d\theta\, .
$$ 
This mean magnetization is shown in the left panel of
Figure~\ref{fig1}.  One can see that unless the pulsar is almost an
orthogonal rotator its value is much higher compared to
$\mean{\sigma}\simeq 10^{-3}$ of the Kennel-Coroniti model and the
values utilized in the 2D numerical simulations, $\mean{\sigma}\simeq
10^{-2}$ \citep{KL03,LDZ04,B05,C09}.  
Unfortunately, $\alpha$ is
poorly constrained from observations.  Using as a guide the value
obtained from fitting the spectrum and pulse profile of the high
energy emission of the Crab pulsar, $\alpha \simeq 45^o$
\citep{HSDF08}, we obtain $\mean{\sigma}\simeq 0.26$.  Thus, the
dissipation of magnetic stripes is apparently unable to resolve the
$\sigma$-problem completely.  
This shortcoming of the striped wind model has already been 
pointed out in \citet{C90}.

The left panel of Figure~\ref{fig1}
shows the distribution of $\sigma$ over the polar angle, where its
value outside of the striped zone is artificially limited by the
rather arbitrary value of $\sim 100$. In reality, this value 
should be determined by the dissipation of fast magnetosonic waves
emitted by the pulsar \citep{L03a}. However, the efficiency of this
emission in 3D numerical simulations of dipolar pulsar magnetospheres
seems to be rather low and thus one would indeed expect a rather high
magnetization in the polar region.

Next we consider the plasma compression at the termination shock of 
such a wind. The magnetic flux conservation ensures that at the
shock $B v_n=$ const, where $v_n$ is the normal component of
velocity. This implies that downstream of the shock the Poynting flux 
is increased by the shock compression factor $\eta=v_{n,1}/v_{n,2}$. 
In the case of strong ultrarelativistic shock,

\be 
\eta(\chi) = 6 \left(1+\chi+\sqrt{1+14\chi+\chi^2} \right)^{-1} \,.  
\ee 
This result holds not only for a perpendicular shock but also
for an oblique shock (see Eq.A14 in \citet{KL11}). Thus, the fraction
of energy injected into PWN in the magnetic form along a wind
streamline is

\be 
\delta_\alpha(\theta) =
\chi_\alpha(\theta)\eta(\chi_\alpha(\theta)) \, .  
\ee 
In the split
monopole model the overall fraction of the wind power injected into
PWN in the magnetic form is given by the integral
 
\be 
\mean{\delta_\alpha}= \frac{3}{2} \int\limits_0^{\pi/2}
\delta_\alpha(\theta)\sin^3\theta d\theta\, .  
\ee 
The function
$\mean{\delta_\alpha}$ is shown in the left panel of Figure~\ref{fig1}
and in Table~\ref{tab1}.  One can see that, unless the magnetic
inclination is close to $90^o$, the fraction of magnetic energy is
quite substantial.  For the guide value of $\alpha \simeq 45^o$
\citep{HSDF08}, we obtain $\mean{\delta}\simeq 0.28$. Thus, almost one
third of the energy supplied into the Crab nebula can be in the
magnetic form.  The middle panel of Figure 1 shows how this flux is
distributed over the polar angle for different magnetic
inclinations. For $\alpha<50^o$ it peaks at the boundary of the
striped zone, but for $\alpha>50^o$ the maximum is inside the striped
zone.

%ttttttttttttttttttttttttttttttttttttttttttttttttttttttttttttttttttttttttt
\begin{table}
\caption{}
   \begin{tabular}{|l|l|l|l|l|l|l|l|l|}
   \hline $\alpha$ & $10^o$ & $20^o$ & $30^o$ & $40^o$ & $50^o$ &
   $60^o$ & $70^o$ & $80^o$ \\ \hline $\mean{\delta_\alpha}$ &
   $\!0.82$ & $\!0.64$ & $\!0.48$ & $\!0.34$ & $\!0.23$ & $\!0.13$ &
   $\!0.061$ & $\!0.017$\\ \hline
   \end{tabular} 
\label{tab1}
\end{table}
%ttttttttttttttttttttttttttttttttttttttttttttttttttttttttttttttttttttttttt

%sssssssssssssssssssssssssssssssssssssssssssssssssssssssssssssssss
\section{Magnetic dissipation inside the nebula} 
\label{sec:dissipation}
%sssssssssssssssssssssssssssssssssssssssssssssssssssssssssssssssss

The observed synchrotron and inverse-Compton emission of the nebula is
well fitted by the ``one-zone'' model with magnetic field of strength
$B\simeq125\,\mu$G \citep{MHZ10}. Although the magnetic field in the
nebula is unlikely to be uniform, this estimate is still more
reliable than the usual equipartition one, which requires an
additional assumption of parity between the energies of magnetic field
and relativistic particles. 
The observed shape of the nebula can be
described as a prolate spheroid with major and minor axes
$a=4.4\,$pc and $b=2.9$ pc \citep{H08}, which gives the volume
$V=(\pi/6)ab^2\simeq 5.7\times10^{56}\mbox{cm}^3$.  The corresponding
total magnetic energy of the nebula is $E_m = 3.5\times10^{47}$erg,
which is significantly below the value of $E_n$ we estimated in
Sec.\ref{sec:supply}.
 
Assuming parity between the energy of relativistic electrons emitting 
synchrotron radiation $E_e$ and the magnetic energy $E_m$, 
\citet{H98} used the observed synchrotron
luminosity of the nebula to derived its equipartition magnetic field
$B_{eq}=330\,\mu$G.  The lower value of $B$ given by \citet{MHZ10}
suggests significant deviation from the energy equipartition. From the
theory of synchrotron emission it follows that the total energy of 
emitting electrons 
$$
   E_e \propto \frac{L_{\rm{syn}}}{B^{3/2}} \left(
   \frac{\int\limits_{\nu_{\rm{min}}}^{\nu_{\rm{max}}} n({\cal E})d\nu }
   {\int\limits_{\nu_{\rm{min}}}^{\nu_{\rm{max}}} n({\cal E})\nu^{1/2}d\nu}\, ,
   \right)
$$
where $L_{\rm{syn}}$ is the total synchrotron luminosity, $n({\cal E})$ is the 
electron energy spectrum, and ${\cal E}\propto (\nu/B)^{1/2}$ is the 
characteristic energy of electron emitting at frequency $\nu$ \citep{P70}. 
For a power-law spectrum the function in the brackets does not depend on $B$ 
and thus $E_e \propto B^{-3/2}$ with sufficient accuracy. 
Since $E_m \propto B^2$ this leads to 

$$ 
E_e = E_m (B/B_{eq})^{-7/2}\, .
$$ 
In the case of the Crab nebula this yields $E_e\simeq 30 E_m \simeq
1.0 \times 10^{49}$erg, which is surprisingly close to our value of
$E_n$, given the simplifications of the model.
 
This result suggests two possible explanations.  First, the energy is 
indeed supplied into the nebula mainly in the form of relativistic
particles.  The analysis presented in the previous section shows that
this would require the Crab pulsar to be almost an orthogonal rotator,
in fact we would need $\alpha\simeq 76^o$.  If however the magnetic
inclination angle is indeed close to $\alpha = 45^o$, obtained in
\citet{HSDF08} via modeling of the pulsed emission, then an efficient
dissipation of magnetic field inside the nebula, accompanied by
particle acceleration, is required to explain the data.

Assuming that a fraction $\mean{\delta}$ of $L_{sp}$ is supplied into
the nebula in the magnetic form, one can find the characteristic
timescale of this dissipation via balancing the supply and dissipation
rates as

\be 
\tau_{md} = \frac{E_m}{\mean{\delta} L_{sp}} \simeq 80
\fracb{\mean{\delta}}{0.3}^{-1} \mbox{yr} .
\label{t-d}
\ee
This is much smaller compared to the dynamical timescale
$\tau_{dn}\simeq 950\,$yr, which justifies the omission of adiabatic
energy losses in this estimate.  Moreover, $\tau_{md}$ exceeds the
light crossing time of the nebula, $\tau_{lc}\simeq12\,$yr, only by a
factor of $\sim 7$. This shows that the magnetic dissipation is a very fast 
process. For example, the speed of magnetic energy supply into 
reconnection zones is likely to be limited from above by $\sim0.1$ of
the Alfv\'en speed \citep{L05}, which in the relativistic MHD is
$$ c_a = c \fracb{\tilde\sigma}{1 +\tilde\sigma}^{1/2},
$$ where $\tilde\sigma = B^2/4\pi w$, where $w=\rho c^2 + \Gamma
p/(\Gamma-1)$ is the relativistic enthalpy and $p$ is the gas
pressure.  In magnetically dominated plasma $\tilde\sigma\gg1$ and
$c_a$ is close to the speed of light, whereas in particle-dominated
plasma with $\tilde\sigma\ll 1$, it can be significantly lower. The
mean $\tilde\sigma$ of the nebula can be estimated as
$<\!\tilde\sigma\!> \simeq 2E_m/E_e \simeq 0.07$ leading to the reconnection
speed $< 0.025c$.  If the reconnection flow had the form of a large-scale 
advection towards the equatorial plane and polar axis, like that proposed 
in \citet{Lt10}, the corresponding dissipation timescale would be  
$\gtrsim 300\,$yr, which seems too high compared to $\tau_{md}$. 
In order to reduce this timescale one would have to involve 
numerous simultaneously active reconnection cites throughout 
the volume of the nebula.

%ssssssssssssssssssssssssssssssssssssssssssssssssssssssss
\section{Discussion}
\label{sec:discussion}
%ssssssssssssssssssssssssssssssssssssssssssssssssssssssss

%%%%%%%%%%%%%%%%%%%%%%
\subsection{Magnetic dissipation and the fine structure of the Crab Nebula}
%%%%%%%%%%%%%%%%%%%%%%

The possibility of efficient magnetic dissipation in the Crab 
nebula raises the question about its observational signatures.   
What kind of structures if any should we expect to see and where?  
Unfortunately, our current understanding of magnetic reconnection is not 
that advanced to make any firm predictions. The most explored and 
firmly associated with magnetic reconnection phenomena in astrophysics are 
solar flares. They involve significant restructuring of magnetic fields 
anchored to the solar surface and distorted by motions in the Sun. 
It is not clear if such grand events may occur under the conditions 
of PWN. Smaller scale ``nanoflares'' could be responsible for heating 
of solar corona and the appearance of bright coronal loops \citep{P72}, 
but even this issue has not been settled yet.  The so-called  
``reconnection exhausts jets'' have been detected in the solar wind via 
in situ measurements using spacecrafts \citep[see][and references wherein]{Gos11}. 
These observations show no evidence of non-thermal particle acceleration or 
electron heating and do not allow to say how far a spacecraft is from a 
reconnection site or even if reconnection is still ongoing at the time of 
observation.  Other examples include Earth's magnetosphere and laboratory 
experiments but these seem to be too specific.

For the purpose of identifying the locations of magnetic dissipation 
in PWN, a non-thermal particle acceleration is its most promising and also 
likely product. This process could brighten up the interfaces between regions 
with different orientation of magnetic field. 
The polarimetric observations of the Crab Nebula 
by \citet{BK91} show that in its central region the degree of polarization
is about 5 times below that for the synchrotron emission in
uniform magnetic field.  They explain this result by the presence
along the line of sight of several cells with randomly oriented
uniform magnetic field. Boundaries of such cells could be the cites 
of ongoing magnetic reconnection and may appear as arcs or filaments  
of enhanced nonthermal emission. 

In fact, the Crab nebula has the most 
spectacular network of optical filaments but they are made of the 
line-emitting thermal plasma of the supernova ejecta ionized by the 
synchrotron radiation of the PWN. 
In the optical continuum, only the bright cores of these filaments can be seen 
and only as absorbing features \citep[e.g.][]{FB90,S98}.  
The radio emission of the Crab nebula has synchrotron origin and given the 
results of the optical continuum observations one would not expect to find the 
filaments in radio images of the nebula. To the contrary, the high  
resolution VLA radio images do show filamentary structure which is as 
spectacular as that of the line emission maps \citep[e.g.][]{BHFB04}.  
Moreover, the radio filaments seem to coincide with the line-emitting 
optical filaments. This was noticed already in the early lower resolution 
study of the nebula with the Cambridge One-Mile radio telescope \citep{W72}. 

A localized enhancement of synchrotron emissivity does not have to be 
related to particle acceleration and may simply reflect a local 
enhancement of magnetic field. Such enhancement could well arise during 
interaction between the high speed flow of relativistic plasma inside the 
PWN with the filaments via the so-called ``magnetic draping'' effect 
\citep[e.g.][]{Lt06}. However, in this case one would generally expect 
the optical non-thermal emissivity to increase as well. Since this is 
not what is observed, other factors should come into play.

The origin of radio emitting electrons (and positrons) of the Crab nebula 
is a long standing mystery. The most natural assumption is that they 
come with the wind from the Crab pulsar just like the higher energy electrons, 
but their number seems to be too high to be accommodated in the current 
models of pair production in pulsar magnetospheres \citep{Ar12}. 
The radio observations of the inner Crab nebula could have settled this 
issue should they revealed the same features associated with the outflow from 
the termination shock as in optics and X-rays, or otherwise.  
Unfortunately, the emerging picture is rather ambiguous. Although radio 
wisps are observed, they do not coincide with the optical ones and 
are noticeably slower \citep{BHFB04}. There is no obvious radio counterpart to the 
optical and X-ray jet either. Given the strong anisotropy of the pulsar 
wind, this may indicate that radio electrons and positrons come from 
different parts of the termination shock. On the other hand, the radio 
wisps could just be some kind of ripples driven by the unsteady outflow
from the termination shock through the PWN. Indeed, the 
MHD simulations show strong convective motion inside the nebula 
which brings plasma
from outer parts of the nebula quite close to termination shock, where 
it is pushed out again by the outflow from the termination shock 
\citep{C09}.

The quantity of radio electrons may be large compared to what is expected 
in the theory of pulsar magnetospheres, but it is tiny compared to what is
available in the line-emitting filaments. It is conceivable that a small 
fraction of the filament plasma becomes lose and mixes with the relativistic 
plasma of PWN. In there, its electrons can be accelerated to relativistic 
energies and produce the observed synchrotron radio emission.    
\cite{KC84b} estimate the energy contained in the radio electrons 
to be of the order of few $\times10^{48}$erg, which is a sizable fraction of 
the total internal energy of the nebula. Thus, in situ acceleration of
radio electrons requires a substantial source of energy. 
This could be the energy of the magnetic field injected into 
the nebula by the pulsar wind, which can indeed be substantial, as  
we argued in Sec.\ref{sec:SW}. The magnetic dissipation can be enhanced
near the line-emitting filaments when magnetic field lines of different 
orientation wrap around the same filament. { The magnetic reconnection 
could also facilitate escape of electrons (and ions) from the filaments into 
Crab's PWN, otherwise suppressed by the low diffusivity across magnetic field 
lines.}  The other possibility is 
the second-order Fermi acceleration by the hydromagnetic turbulence 
driven by various instabilities \citep{B11}. { The existence of two 
synchrotron components of different origin is supported by the combined 
radio and mm-wavelength observations \citep{BNC02}. The data suggest low 
energy cutoff around 100~GHz in the emission of the electrons supplied 
by the termination shock, thus supporting a different origin of radio 
emitting electrons. However, the relatively smooth matching of radio and 
optical/infrared components of the integral spectrum may be problematic for 
this model, particularly when seen in a number of PWN \citep{B11}.}    

\citet{VRV92} state that the VLA images also show filaments which do not
have line emitting counterparts. The spectral data do not show 
any noticeable variation of the radio spectral index across the filaments 
\citep{B97}. Since the synchrotron life-time of radio emitting electrons
significantly exceeds the age of the nebula, this is not very surprising.    
Some of the filaments are also seen in the X-ray band \citep{STF06}. 
Since the life-time of X-ray emitting electrons is quite short, one 
would expect to see hardening of the X-ray spectrum of the filaments compared 
to the diffuse background if these filaments were indeed the 
cites of particle acceleration. However, the observations give no 
evidence of such hardening and the photon index is very soft, 
$\alpha\sim3-4$ \citep{STF06}, creating a problem for any model where 
these features act as acceleration cites of electrons emitting 
in X-rays \citep{STF06}.     

The continuum optical images of the Crab nebula reveal fine 
fibrous structure somewhat reminiscent of solar coronal loops 
\citep{FB90,H95}. However, it is not clear if this is a product of 
``nanoflares'' or simply reflects inhomogeneous structure of magnetic 
field.

%%%%%%%%%%%%%%%%%%%%%%
\subsection{Implications for numerical simulations of PWN}
%%%%%%%%%%%%%%%%%%%%%%

The 2D RMHD numerical simulations of the Crab nebula
\citep{KL03,LDZ04,B05,C09} have been very successful in reproducing
many key properties of the nebula, such as its jet-torus, the
brightness asymmetry, wisps, and even the bright ``inner knot''
\citep{H95}. In agreement with the observations, the proper motion of
jets and wisps produced in the simulations is relatively low, $v=0.2-0.7c$, 
as expected downstream of an almost purely hydrodynamical shock
wave. This success leaves little doubt that the numerical models capture 
the physics of the nebula quite well.

However, the overall low wind magnetization
utilized in these models, $<\!\sigma\!>\simeq 10^{-2}$, is in conflict
with what we would expect in the striped wind model without imposing
very large magnetic inclination angle of the pulsar. 
This choice of $\sigma$ has been influenced by the very low
value required in the Kennel-Coroniti model in order to have a
termination shock in their 1D solution. However, the flow dynamics of
the 2D numerical solutions is already very different, as it involves
large scale circulation and mixing.  Although \citet{KL04} did find that, 
in qualitative agreement with predictions of the Kennel-Coroniti model, 
the size of the termination shock decreased with $<\!\sigma\!>$, no attempts 
have been made to study models with  $\sigma\gg 10^{-2}$. As the result,  
one cannot claim yet that 2D numerical simulations rule them out.  
As the shock size is
determined by the balance between the wind ram pressure and the total
pressure in the nebula, this tendency can be explained by the
stronger axial compression of the nebula by the magnetic hoop
stress in models with higher $\sigma$. 
However, this compression is certainly excessive in 2D
models, being enforced by the condition of axial symmetry which does
not allow development of the kink instability \citep{B98}. 
The 3D numerical study of z-pinch configurations by \citet{M09,M11} 
confirms this expectation. Thus, the ultimate
answer to the question whether $<\!\sigma\!>\,\,\gg 10^{-2}$ is 
allowed by the RMHD model will only be found in future 3D simulations 
of PWN.

If at high latitudes the pulsar wind is free from stripes and has 
high $\sigma$ then downstream of the termination shock
one would expect a very fast flow, with the Lorentz factor
$\gamma\sim\sigma^{1/2}$ in the case of perpendicular shock
\citep{KC84a} and even higher in the case of oblique shock
\citep{KL11}. Downstream of a perpendicular shock the flow is subsonic
(or sub-fast-magnetosonic to be more precise) and can smoothly
decelerate down to $\gamma\simeq 1$ inside the nebula.  Downstream of
an oblique shock it may remain supersonic and a secondary shock will
have to appear somewhere on its way.  So far the observations of the
Crab nebula show no evidence of such a secondary shock or such a fast
flow.  This may well be related to the low dissipation efficiency of
shocks in highly magnetized plasma \citep[e.g.][]{KC84a,K12}, as well
as the inability of such shocks to accelerate non-thermal particles
\citep{SS09,SS11a}.  Further investigation is required to clarify this
issue.

%%%%%%%%%%%%%%%%%%%%%%
\subsection{Magnetic dissipation and Crab's gamma-ray flares}
%%%%%%%%%%%%%%%%%%%%%%

The recently discovered strong flares of gamma-ray emission from the 
Crab nebula at the energies $\sim 1\,$GeV with duration about few days 
\citep{T11,A11} could be very important for understanding the physics 
of highly magnetized relativistic plasma. \citet{KL11} argued that 
the gamma-rays of these energies could originate from the most compact 
known bright feature of the Crab nebula, the so-called ``inner knot'', 
which they explain as a Doppler-boosted emission from the termination 
shock. However, their model predicts synchronous variability of the knot 
emission in gamma-rays and optics, which does not seem to be the 
case \citep{Ar12}. The only other promising alternative seem to be 
explosive magnetic reconnection.      

However, the properties of these flares suggest that they may not 
be representative of the energetically dominant magnetic dissipation 
process in the nebula.     
First, the dissipation time scale
given by Eq.\ref{t-d} is at least three orders of magnitude longer
than the typical flare duration.  It is possible that the tearing
instability produces much smaller structures inside the large scale
current sheets, however in this case one would expect a whole spectrum
of time scales to be present.  
Second, the statistical model of flares by \citep{CL12} gives the total 
energy release rate which is three orders of magnitude below the spindown 
power of the Crab pulsar and hence significantly lower than the 
magnetic dissipation rate given by Eq.\ref{t-d}.    
Third, the current reconnection models of these flares involve strong 
magnetic fields, of order $1000\,\mu$G \citep{UCB11,CUB12}, 
and/or large bulk Lorentz factors $\Gamma\gtrsim\mbox{few}$ 
\citep{KL11,CL12}.  Such conditions are not typical for the Crab nebula. 
Finally, so far the flares have not been identified with any particular 
kind of events seen at other energies.  

Given the required conditions for the flares, their
most likely location is the polar region near 
the termination shock, where the freshly supplied plasma can have 
very high magnetization and streams with ultra-relativistic speeds\footnote{ 
A similar conclusion was reached recently by Y.Lyubarsky at a conference 
presentation (http://www.iasf-roma.inaf.it/Flaring\_Crab/index.html).}.  
Large Lorentz factors could also be
produced during fast reconnection events inside high-$\tilde\sigma$
plasma, which again points out towards the inner polar region of the
Crab nebula, where the observations reveal the Crab jet. 
High magnetization also implies Alfv\'en speed approaching the speed of 
light and hence the fastest possible magnetic reconnection speed.
\citet{CUB12} also point out that magnetic field in this region 
can be much stronger than on average due to the strong axial compression 
associated with the z-pinch. The region at the base of the Crab jet, 
the so-called ``anvil'', is in fact the most active region in the nebula 
\citep{H02}.

%ssssssssssssssssssssssssssssssssssssssssssssssssssssssss
\section{Summary}
\label{sec:conclusions}
%ssssssssssssssssssssssssssssssssssssssssssssssssssssssss

\begin{enumerate} 

\item 
We have calculated the power of high-$\sigma$ striped pulsar wind
which remains as the Poynting flux after total dissipation of its
stripes in the split-monopole approximation.  
The results show that the pulsar has to be an almost exact
orthogonal rotator for the mean wind $\sigma$ to reduce down to the
very low values suggested by the Kennel-Coroniti model (and to some degree 
by the current axisymmetric numerical models of the Crab nebula).  For the
more realistic magnetic inclination angle $\alpha\simeq45^o$, about 30
percent of the wind power is retained in the form of the Poynting
flux. While low magnetization is achieved in the equatorial plane, in
the polar zone the magnetization remains very high.
 
\item 
Given the relatively long spindown time of the Crab pulsar and low
radiative losses, we find that out of $E \simeq 3.7\times
10^{49}$erg of energy that has been supplied by the pulsar wind into the
nebula $E_n \simeq 1.3\times10^{49}$erg should still remain as its
internal energy, sheared between magnetic field and relativistic
particles.

\item 
The observations of synchrotron and inverse-Compton emission of
the Crab nebula indicate that most of $E_n$ is stored in relativistic
electrons and positrons, and only $E_m\simeq 3.5\times 10^{47}$erg in
the magnetic field.  This may be simply down to the fact that from the
start the energy is injected into the nebula mostly in the form of
relativistic particles.  In the striped wind model, this would imply that
the Crab pulsar is almost an exact orthogonal rotator. Alternatively,
most of the injected magnetic energy may have been dissipated and
transferred to the particles via magnetic reconnection events.

\item 
Using the magnetic inclination angle of the Crab pulsar derived from
modeling of its high energy pulsed emission, $\alpha=45^o$, we
estimate the characteristic timescale of magnetic dissipation in the
Crab nebula to be $\tau_{md}\sim 80\,$yr.  This relatively short
timescale implies complex structure in the magnetic field distribution
inside the nebula, which is supported by the radio and optical
observations.

\item
Since the scale of deduced magnetic dissipation inside the Crab nebula
strongly depends on the magnetic inclination angle of its pulsar, accurate
observational measurements of this angle would be very important. 
A search for signs of ongoing magnetic dissipation, 
such as particle acceleration inside the nebula, is another important 
direction of observational studies. It seems plausible that the observed 
enhanced radio emissivity in vicinity of line-emitting filaments is 
a result of magnetic dissipation.  

\item 
The recently discovered gamma-ray flares may be the first strong indication 
of magnetic reconnection inside the Crab nebula. However, their short 
timescale, low energetics, and extreme conditions requires in the 
current theoretical models suggest that these events may not be 
representative of the dominant magnetic dissipation  process in the 
nebula. 
\end{enumerate}

%%%%%%%%%%%%%%%%%%%%%%%%%%%%%%%%%%%%%%%%%%%%%%%%%%%%%%%%%%%%%%%%%
\section*{Acknowledgments}
%%%%%%%%%%%%%%%%%%%%%%%%%%%%%%%%%%%%%%%%%%%%%%%%%%%%%%%%%%%%%%%%%
We thank Y.Lyubarsky for careful reading of the manuscript and finding 
an error in the original derivations as well as the anonymous referee 
for numerous suggestions on improving the presentation.  
This research was funded by STFC under the standard grant ST/I001816/1.

%%%%%%%%%%%%%%%%%%%%%%%%%%%%%%%%%%%%%%%%%%%%%%%%%%%%%%%%%%%%%%%%%

\end{document}